\documentclass[12pt]{article}
\usepackage{a4wide}

\usepackage{latexsym}
\usepackage{amsmath}
\usepackage{amsfonts}

\usepackage{cite}
\usepackage{pslatex}
\usepackage[latin1]{inputenc}
\usepackage[T1]{fontenc}

\newcommand{\bq}{\begin{eqnarray}}
\newcommand{\eq}{\end{eqnarray}}

\newcommand{\eps}{\varepsilon}

\begin{document}

\thispagestyle{empty}

\begin{flushright}
 MZ-TH/08-06 \\
\end{flushright}

\vspace{1.5cm}

\begin{center}
  {\Large\bf Symmetry breaking of gauge theories down to Abelian sub-groups\\
  }
  \vspace{1cm}
  {\large Stefan Weinzierl\\
  \vspace{1cm}
      {\small \em Institut f{\"u}r Physik, Universit{\"a}t Mainz,}\\
      {\small \em D - 55099 Mainz, Germany}\\
  } 
\end{center}

\vspace{2cm}

\begin{abstract}\noindent
  { 
I re-derive the lowest order effective Lagrangian for electro-weak symmetry breaking 
without the use of Goldstone's theorem for spontaneously broken global symmetries
and without the assumption of a custodial symmetry.
I consider the breaking of a local symmetry with gauge group $G$ down to an Abelian sub-group $K$
and construct a gauge-invariant functional with one free parameter $v$, such that $v=0$ 
corresponds to a gauge theory with gauge group $G$, while $v \rightarrow \infty$ corresponds to a gauge
theory with gauge group $K$.
  }
\end{abstract}

\vspace{1cm}


\vspace*{\fill}

\newpage
\section{Introduction}
\label{sect:intro}

The standard mechanism for the generation of masses for the electro-weak gauge bosons is the
Higgs mechanism \cite{Higgs:1964ia,Higgs:1964pj,Higgs:1966ev,Englert:1964et,Guralnik:1964eu,Kibble:1967sv}.
It predicts an additional spin zero particle, the Higgs boson. From direct searches we know 
that it must be heavier than $114 \;\mbox{GeV}$. On the other hand, electro-weak precision
measurements prefer a value below this limit.
It is therefore legitimate to investigate alternatives to the Higgs mechanism.

An alternative is an approach based on an effective theory, which would
just add the three required pseudo-Goldstone fields, 
but no Higgs field.
These effective Lagrangians are usually derived by assuming a global symmetry which is spontaneously broken.
By Goldstone's theorem \cite{Goldstone:1961eq,Goldstone:1962es} there will be a massless scalar field
for every broken generator of a global symmetry.
In a second step these models are gauged. This converts the Goldstone fields into pseudo-Goldstone fields,
which provide the longitudinal degrees of freedom for the massive electro-weak gauge bosons \cite{Cornwall:1974km,Lee:1977eg,Chanowitz:1985hj}.

Chanowitz, Golden and Georgi \cite{Chanowitz:1986hu,Chanowitz:1987vj}
have shown that for the breaking of the global symmetry there are precisely two possibilities:
$SU_L(2)\times SU_R(2) \rightarrow SU_{L+R}(2)$ and
$SU_L(2)\times U_Y(1) \rightarrow U_{QED}(1)$.
Due to its similarity with chiral perturbation theory the first possibility is often called
chiral electro-weak symmetry breaking \cite{Appelquist:1980vg,Longhitano:1980iz,Longhitano:1980tm}
and has been discussed extensively in the literature
\cite{Donoghue:1989qw,Dawson:1990cc,Dobado:1989ax,Dobado:1990zh,Espriu:1991vm}.
The un-broken global symmetry $SU_{L+R}(2)$ is usually called the custodial symmetry \cite{Sikivie:1980hm}.
Upon gauging the model the custodial symmetry is explicitly broken.
Chiral electro-weak symmetry breaking predicts the tree-level value $\rho=1$ for the $\rho$-parameter \cite{Veltman:1976rt,Veltman:1977kh}.

The second symmetry breaking pattern $SU_L(2)\times U_Y(1) \rightarrow U_{QED}(1)$
has been discussed in \cite{Chanowitz:1986hu,Chanowitz:1987vj}
and leaves the tree-level value of the $\rho$-parameter unconstrained.

From experimental measurements we know that $\rho$ is very close to $1$. The experiments would therefore point towards
chiral electro-weak symmetry breaking if they would have to decide among the two options.
However, there are several questions which can be raised.
First of all it is not clear why one should start from a global symmetry.
In the electro-weak case we are interested in the breaking of a local symmetry, not a global one.
Secondly in the case of chiral electro-weak symmetry breaking the global symmetry group $SU_L(2) \times SU_R(2)$ is
not identical with the gauge group $SU_L(2) \times U_Y(1)$.
The identification of $U_Y(1)$ with a one-parameter sub-group of $SU_R(2)$ is problematic.
In fact, the quantum numbers of the fermions suggest, that the $U_Y(1)$ is itself the result
of a symmetry breaking
$U_{B-L}(1) \times SU_R(2) \rightarrow U_Y(1)$, where $B-L$ stands for baryon number minus lepton
number.
Therefore the generator of the hyper-charge would contain a term, which can be identified
with the third generator of $SU_R(2)$. In addition, there would be a second term, corresponding
to the generator of $U_{B-L}(1)$, and commuting with $SU_L(2)$ and $SU_R(2)$.

It is therefore interesting to ask what assumptions are really needed to derive the lowest order
effective Lagrangian for electro-weak symmetry breaking with $\rho=1$.
In this letter I present a derivation of the effective Lagrangian which tries to keep 
the necessary assumptions to a minimum.
The derivation does not make use of Goldstone's theorem for spontaneously broken global symmetries
nor does it assume a custodial symmetry.
I will treat directly the breaking of a local gauge symmetry with gauge group $G$ down to a gauge group $K$.
For the application towards electro-weak theory this is the breaking of a local
$SU_L(2) \times U_Y(1)$ symmetry down to a local $U_{QED}(1)$ symmetry.
I will assume the following:
\begin{description}
\item{(i)} The un-broken sub-group is Abelian.
\item{(ii)} There exists a decomposition of the Lie algebra of $G$ as a vector space
into the directions of $K$ and $G/K$.
It is not assumed that this decomposition is orthogonal with respect to the inner
product of the Lie algebra.
This decomposition can be given in the form of a projection onto the directions of $G/K$.
Of particular interest are cases where this decomposition is such that the coset space $G/K$ 
is isomorphic to a group.
\end{description}
While assumption (i) clearly is justified for the case of interest $SU_L(2) \times U_Y(1) \rightarrow U_{QED}(1)$,
it will turn out that assumption (ii) is essential to establish $\rho=1$.
Different decompositions correspond to different values of the $\rho$-parameter.
$\rho = 1$ corresponds to a decomposition where $G/K$ is isomorphic to $SU_L(2)$.
To derive the effective Lagrangian I construct a gauge-invariant functional
with one free parameter $v$, such that $v=0$ 
corresponds to a gauge theory with gauge group $G$, while $v \rightarrow \infty$ corresponds to a gauge
theory with gauge group $K$.
The functional involves an integration over all gauge transformations.
For infinitesimal gauge transformations, the integration over the gauge transformations
of the un-broken sub-group factorise, leaving an integration over the moduli space.
This integration is identified with the integration over the pseudo-Goldstone fields.
When the functional is expanded in the pseudo-Goldstone fields, the first term yields
the standard term needed to generate the masses of the electro-weak gauge bosons.

In this letter I focus on the lowest order effective Lagrangian. As usual in effective theories,
terms corresponding to higher-dimensional operators have to be added.

This letter is organised as follows: The next section introduces the notation.
Sections~\ref{sect:non-linear} and \ref{sect:chiral} review non-linear realisations of a group 
and chiral electro-weak symmetry breaking.
In section~\ref{sect:derivation} the functional for the symmetry breaking of a local symmetry down
to a local Abelian symmetry is derived.
Section~\ref{sect:application} applies this formalism to the electro-weak theory.
Finally section~\ref{sect:conclusions} contains the conclusions.

\section{Notation}
\label{sect:notation}

Let $G$ be a compact connected Lie group and let $K$ be an Abelian continuous sub-group of $G$. 
I denote the dimension of $G$ by $N$ and the dimension of $K$ by $n$.
The Lie algebra of $G$ is denoted by $\mathfrak g$, the one of $K$ is denoted by $\mathfrak k$.
As a vector space we can decompose $\mathfrak g$ into $\mathfrak k$
and the sub-space $\mathfrak b$ generated by the broken generators:
\bq
 {\mathfrak g} & = & {\mathfrak k} + {\mathfrak b}.
\eq
I denote the projections onto the individual sub-spaces by
\bq
 P_{\mathfrak k}: & & \mbox{projection onto } {\mathfrak k},
 \nonumber \\
 P_{\mathfrak b}: & & \mbox{projection onto } {\mathfrak b}.
\eq
The generators of the group are normalised according to
\bq
 \mbox{Tr}\; \left( T^\dagger T \right) & = & \frac{1}{2}.
\eq
I will use the notation $T^a$ for a generic base.
In this letter I do not assume that the generators are orthogonal.
In particular while I will choose an orthogonal base for $\mathfrak k$
as well as an orthogonal base for the vector space $\mathfrak b$,
I do not assume that the vector spaces $\mathfrak k$ and $\mathfrak b$ are orthogonal.
It will be convenient to work in a specific base, which is a Cartan base.
A Cartan base contains the maximal number of mutually commuting generators. 
The number of mutually commuting generators equals the rank of the Lie algebra and is denoted by $r$.
I will label the mutually commuting generators $H^a$ and the remaining generators $E^a$. 
They can chosen to satisfy
\bq
 \left[ H^a, H^b \right] = 0,
 & &
 \left[ H^a, E^b \right] = \alpha_a^{(b)} E^b.
\eq
In the last equation no summation over $b$ is implied.
The vector $\alpha^{(b)}=(\alpha_1^{(b)}, ..., \alpha_r^{(b)} )$ is called the root vector of the generator $E^b$.
A basic theorem on Lie algebras states that for any generator $E$ with non-zero root vector $\alpha$ there is 
another generator with root vector $-\alpha$.
The generators $E^a$ therefore come always in pairs and it will be convenient to label them
$E^a$ and $E^{-a}$, where $a$ takes the values $1,...,(N-r)/2$.
If $K$ is Abelian, we can choose a Cartan base such that
$H^1, ..., H^n$ are the generators of $K$.
In this case we can further decompose ${\mathfrak b}$ as a vector space into
\bq
 {\mathfrak b} & = & {\mathfrak h} + {\mathfrak e},
\eq
where $\mathfrak h$ is generated by the remaining mutually commuting generators
$H^{n+1}, ..., H^r$, and $\mathfrak e$ is generated by 
$E^{1}, E^{-1}, ..., E^{(N-r)/2}, E^{-(N-r)/2}$.
As a vector space we have therefore the decomposition
\bq
 {\mathfrak g} & = & {\mathfrak k} + {\mathfrak h} + {\mathfrak e}.
\eq
Let us further agree that if we label the generators by $T^a$ with $a=1,...,N$, then we assume
that they are ordered such that the first $n$ generators correspond to $\mathfrak k$, the next
$(r-n)$ generators correspond to $\mathfrak h$ and the remaining $(N-r)$ generators correspond to
$\mathfrak e$.

Let us now consider a gauge theory with gauge group $G$.
The gauge potential and the field strength are denoted by $A$ and $F$, respectively:
\bq
 A = \frac{g}{i} T^a A^a_\mu dx^\mu,
 & &
 F = \frac{g}{2i} T^a F^a_{\mu\nu} dx^\mu \wedge dx^\nu.
\eq
$A$ is a one-form which takes values in the Lie algebra $\mathfrak g$, $F$ is a two-form which also takes values in
$\mathfrak g$. 
The coupling constant is denoted by $g$.
As one frequently encounters differential forms which take values in the Lie algebra $\mathfrak g$, I adopt the 
convention that for an $k$-form $\omega$
\bq
 || \omega || & = & \mbox{Tr}\; \left( \omega \wedge \ast \omega \right).
\eq
$\ast \omega$ is the Hodge-dual of $\omega$, defined on a $D$-dimensional flat manifold by
\bq
 \ast \left( T^a \omega^a_{\mu_1 ... \mu_k} dx^{\mu_1} \wedge ... \wedge dx^{\mu_k} \right)
 & = & 
 \frac{1}{(D-k)!}
 \left( T^a \omega^{a \mu_1 ... \mu_k} \right)^\dagger
 \eps_{\mu_1 ... \mu_k \mu_{k+1} ... \mu_D} dx^{\mu_{k+1}} \wedge ... \wedge dx^{\mu_D}.
\;\;\;
\eq
We have for example
\bq
 || F || = - \frac{g^2}{4} F^a_{\mu\nu} F^{a\mu\nu} d^4x,
 & &
 || A || = - \frac{g^2}{2} A^a_{\mu} A^{a\mu} d^4x.
 \nonumber
\eq
The minus sign is related to the fact, that in Minkowski space
the contraction of the total anti-symmetric tensor yields
\bq
 \eps_{\mu\nu\rho\sigma} \eps^{\mu\nu\rho\sigma} & = & -24.
\eq

\section{Non-linear realisations}
\label{sect:non-linear}

In this section I review the construction of Coleman, Wess and Zumino \cite{Coleman:1969sm}
for non-linear realisations of a group $G$.
In this section I do not assume that the sub-group $K$ is Abelian.
As mentioned in the previous section I assume that
the generators $T^a$ are ordered, such that the first $n$ generators correspond to the un-broken
sub-group $K$, the remaining $N-n$ generators are then the broken generators.
An element of the coset space $G/K$ can be parameterised as
\bq
\label{non-linear}
 U & = & \exp \left( i \sum\limits_{a=n+1}^N T^a \chi^a \right) 
         \exp \left( i \sum\limits_{b=1}^n T^b \xi^b \right).
\eq
The standard choice for a coset representative is $\xi^b=0$ for $b=1,...,n$.
An element $U_1$ of $G$
\bq
 U_1 & = & \exp \left( i \sum\limits_{a=1}^N T^a \theta^a \right)
\eq
acts on $U$ from the left.
The result can again be brought in the form of eq.~(\ref{non-linear}):
\bq
 U_1 U & = &
 \exp \left( i \sum\limits_{a=n+1}^N T^a {\chi'}^a \right) 
 \exp \left( i \sum\limits_{b=1}^n T^b {\xi'}^b \right).
 \;\;\;\;
\eq
The new coordinates ${\chi'}^a$ and ${\xi'}^b$ depend on $\chi$ and $\theta$.
In general we have ${\xi'}^b \neq 0$, and the second exponential can be thought of as 
a compensating function needed to return to the given choice of coset representative $\xi^b=0$.
In the case were the group $G$ admits an automorphism $R : G \rightarrow G$ such that
\bq
\label{automorphism}
 T^a & \rightarrow & 
 \left\{ \begin{array}{rl}
 T^a, & a=1,...,n, \\
 -T^a, & a=n+1,...,N, \\
 \end{array} \right.
\eq
the compensating function can be eliminated by considering the transformation
\bq
\label{no_comp_fct}
 U_1 \exp \left( 2 i \sum\limits_{a=n+1}^N T^a \chi^a \right) R\left(U_1^{-1}\right)
 & = &
 \exp \left( 2 i \sum\limits_{a=n+1}^N T^a {\chi'}^a \right).
\eq
Such an automorphism $R$ exists for the breaking of $SU_L(N) \times SU_R(N)$ down to the 
diagonal sub-group $SU_{L+R}(N)$ for all $N$.

In this letter we are interested in a slightly more general case, 
namely where the compensating function -- although it cannot be eliminated -- is still independent of 
the coset coordinates $\chi$.
In this case
the transformed coordinates ${\xi '}$ depend on ${\xi}$, but not on
$\chi$.
In other words, the compensating function is constant on the coset space $G/K$.
Let us give a concrete example for this case. Assume that
\bq 
 G & = & G' \times U_{Y_1} \times ... \times U_{Y_n},
 \nonumber \\
 K & = & U_{Q_1} \times ... \times U_{Q_n},
\eq
where $G'$ is a semi-simple Lie group of rank $r$ with $r \ge n$. The diagonal generators of $G'$ are denoted
by ${H'}^a$ with $a=1,...,r$.
The generators $Q^j$ of $U_{Q_j}$ are assumed to be linear combinations of the generators $Y^j$ of $U_{Y_j}$ 
and the diagonal generators ${H'}^a$, say
\bq
 Q^j & = & \frac{1}{\sqrt{2}} \left( Y^j + {H'}^j \right).
\eq
The coset space $G/K$ is isomorphic to $G'$
and we can take as coset representative
\bq
 U & = & \exp \left( i \sum\limits_{a=1}^{N-n} {T'}^a \chi^a \right), 
\eq
where ${T'}^a$ denote the generators of $G'$.
Note that the set $\{{T'}^a,Q^j\}$ is a non-orthogonal base of the Lie algebra of $G$.
For an element
\bq
 U_1 & = & \exp \left( i \sum\limits_{a=1}^{N-n} {T'}^a \theta^a + i \sum\limits_{j=1}^n Y^j \tilde{\theta}^j \right)
\eq
of $G$ acting from left on $U$ we have
\bq
 U_1 U & = & U' V,
\eq
where $V \in K$ and
\bq
 U' & = & 
    \exp \left( i \sum\limits_{a=1}^{N-n} {T'}^a \theta^a \right) 
    U 
    \exp \left( - i \sqrt{2} \sum\limits_{j=1}^n {H'}^j \tilde{\theta}^j \right),
 \nonumber \\
 V & = & \exp \left( i \sqrt{2} \sum\limits_{j=1}^n Q^j \tilde{\theta}^j \right).
\eq
Let us remark that since we assumed that the rank $r$ of $G'$ satisfies $r \ge n$, we could always
embed the torus $U_{Y_1} \times ... \times U_{Y_n}$ in a second copy of $G'$ by identifying $Y^j$ with ${H'}^j$:
\bq
 G' \times U_{Y_1} \times ... \times U_{Y_n} & \subset & G' \times G'.
\eq
However, this enlargement of the group $G$ is an assumption I do not want to make in this letter.

\section{Chiral electro-weak symmetry breaking}
\label{sect:chiral}

In this section I review the standard derivation of the effective Lagrangian
for chiral electro-weak symmetry breaking \cite{Appelquist:1980vg,Longhitano:1980iz,Longhitano:1980tm}.
To describe chiral electro-weak symmetry breaking, let us start from 
the Lagrangian of the Higgs sector without any coupling to gauge fields:
\bq
\label{lagrangian_higgs}
{\cal L}_{Higgs} & = & \left( \partial_\mu \phi \right)^\dagger \left( \partial^\mu \phi \right) 
 + \mu^2 \phi^\dagger \phi - \frac{1}{4} \lambda \left( \phi^\dagger \phi \right)^2.
\eq
We set now
\bq 
 \phi & = & \frac{1}{\sqrt{2}} 
 \left( \begin{array}{c}
   \chi^1 - i \chi^2 \\
   \sigma + i \chi^3 \\
 \end{array} \right),
 \nonumber \\
 \Sigma & = & \frac{1}{2} \sigma + i \chi^a \frac{1}{2} \sigma^a,
\eq
where $\sigma^a$ denotes the Pauli matrices and $\sigma$ a scalar field.
In the Higgs model we usually set $\sigma = v + H$.
The Lagrangian (\ref{lagrangian_higgs}) can be written as 
\bq
\label{lagrangian_higgs2}
{\cal L}_{Higgs} & = & 
 \; \mbox{Tr}\; \left( \partial_\mu \Sigma \right)^\dagger \partial^\mu \Sigma
 -
 \frac{\mu^2}{v^2} \left( \mbox{Tr}\; \left( \Sigma \right)^\dagger \Sigma - \frac{1}{2} v^2 \right)^2
 +
 \frac{1}{4} \mu^2 v^2.
 \nonumber 
\eq
Eq.~(\ref{lagrangian_higgs2}) is invariant under a global $SU_L(2) \times SU_R(2)$ symmetry.
In the limit of a heavy Higgs mass $m_H = \sqrt{2} \mu \rightarrow \infty$ the second term enforces
\bq
 \mbox{Tr}\; \left( \Sigma \right)^\dagger \Sigma & = & \frac{1}{2} v^2.
\eq
It follows that in this limit
\bq
 U & = & \frac{2}{v} \Sigma
\eq
is an element of $SU(2)$.
The Lagrangian
\bq
\label{lagrangian_chiral}
{\cal L}_{chiral} & = &
 \frac{v^2}{4}
 \; \mbox{Tr}\; \left( \partial_\mu U \right)^\dagger \partial^\mu U,
\eq
is the lowest order Lagrangian for the breaking of a global $SU_L(2) \times SU_R(2)$ symmetry 
down to $SU_{L+R}(2)$.
The Lagrangian in eq.~(\ref{lagrangian_chiral}) is invariant under a global $SU_L(2) \times SU_R(2)$ symmetry.
In order to add the couplings to the gauge fields, one replaces in eq.~(\ref{lagrangian_chiral}) 
the partial derivatives
by covariant ones. One arrives at the
lowest order Lagrangian for chiral electro-weak symmetry breaking 
\bq
\label{Lagrangian_chiralSB}
{\cal L}_{\chi S B} & = &
 \frac{v^2}{4}
 \; \mbox{Tr}\; \left( D_\mu U \right)^\dagger D^\mu U,
\eq
where the covariant derivative acts on $U$ as follows:
\bq
\label{def_covariant_derivative_chiral}
 D_\mu U & = & 
 \partial_\mu U
 - i g W^a_\mu \frac{1}{2} \sigma^a U
 + i g' U B_\mu \frac{1}{2} \sigma^3.
\eq
The Lagrangian in eq.~(\ref{Lagrangian_chiralSB}) is invariant under local $SU_L(2) \times U_Y(1)$ transformations.
However, the symmetry under global $SU_R(2)$ transformations is lost due to the presence
of $\sigma^3$ in the $U_Y(1)$-part.

Let us summarise the basic assumptions of chiral electro-weak symmetry breaking:
One assumes a global $SU_L(2) \times SU_R(2)$ symmetry, which is spontaneously broken
down to $SU_{L+R}(2)$.
The global symmetry is partially made local by gauging $SU_L(2)$ and a one-parameter sub-group of $SU_R(2)$.
This one-parameter sub-group is identified with $U_Y(1)$.
Gauging just a one-parameter sub-group of $SU_R(2)$ destroys the global $SU_R(2)$ symmetry.

In the next section I will describe a formalism to derive the effective Lagrangian in eq.~(\ref{Lagrangian_chiralSB}) 
without considering first global symmetries and without the assumption of an additional $SU_R(2)$ symmetry.

\section{The functional for symmetry breaking}
\label{sect:derivation}

In this section I derive a gauge-invariant functional for symmetry breaking.
I first consider gauge-equivalent configurations 
in section~\ref{subsect:equivalent}.
The factorisation for infinitesimal gauge transformations is discussed in section~\ref{subsect:infinitessimal}.
Effects for finite gauge transformations due to the measure are discussed in section~\ref{subsect:finite}.

\subsection{Gauge-equivalent configurations}
\label{subsect:equivalent}

Let us consider a Yang-Mills theory where the symmetry is broken from the gauge group $G$ 
down to an Abelian sub-group $K$.
I shall introduce a parameter $v$ (with the dimension of a mass), such that $v = 0$ corresponds to the unbroken
theory with gauge group $G$ and that $v \rightarrow \infty$ corresponds to a theory with gauge group $K$.
Let us first consider the latter case of a Yang-Mills theory with gauge group $K$. We can embed $K$ in $G$
and with our ordering of the generators we have
\bq
\label{condition_H}
 A^a_\mu & = & \left\{ 
 \begin{array}{cll}
 A^a_\mu, & & a=1,...,n, \\
 0, & & a=n+1,...,N.
 \end{array}
 \right.
\eq
The components in the direction of ${\mathfrak b}$ are simply zero.
This is left un-changed by any gauge transformation of $K$.
However, a gauge transformation in the full group $G$ rotates in general the gauge potential in the directions
of ${\mathfrak b}$.
Let us now look at a general gauge potential $A$ for the gauge group $G$.
We can now ask under which conditions is this gauge potential equivalent to the one in eq.~(\ref{condition_H}).
This is the case, if we can find a gauge transformation $U$ in $G$, such that $A$ can brought in the form
of eq.~(\ref{condition_H}). 
With the help of the projection $P_{\mathfrak b}$
this can be formulated as follows: 
\bq
\label{condition_projection}
 P_{\mathfrak b} \left( A^U \right) & = & 0,
\eq
where $A^U$ denotes the gauge transform of $A$ by $U$:
\bq
 A^U & = & U^{-1} A U + U^{-1} d U.
\eq
Such a $U$ is not unique, we still have the freedom to perform gauge transformations in $K$.
If $V$ is a gauge transformation of $K$, then also
\bq
 U' & = & U V
\eq
is a solution to eq.~(\ref{condition_projection}).
Our aim is now to construct a gauge-invariant functional, which disfavours configurations, which are not of the form 
as in eq.~(\ref{condition_H}).
Let us consider the functional
\bq
\label{def_Z3}
 Z[A] & = & 
 \int {\cal D}U \exp\left( - i \frac{v^2}{4} \int || P_{\mathfrak b}(A^U) || \right).
\eq
The integration is over all gauge transformations $U$ of $G$.
The functional $Z$ depends on the gauge potential $A$
and is a measure how far away a configuration is from the form of eq.~(\ref{condition_H}).
$v$ is an arbitrary constant with the dimension of a mass.
In the functional integral over $A$ we will now weight every configuration with the factor
(\ref{def_Z3}).
For $v=0$ the factor is unity and each configuration receives the same weight.
This corresponds to the un-broken phase.
On the other hand $v \rightarrow \infty$ will enforce eq.~(\ref{condition_H}).

The functional $Z[A]$ is invariant under all gauge transformations $U_1$ of $G$:
\bq
\label{invariance_G}
 Z\left[A^{U_1}\right] & = & Z[A].
\eq
This is easily verified with the help of
\bq
 \left( A^{U_1} \right)^U & = & A^{U_1 U}
\eq
and the fact that the measure is invariant:
\bq
 {\cal D}U & = & {\cal D}\left(U_1 U\right).
\eq
The functional $Z[A]$ involves an  integration over all gauge transformations $U$ of $G$.
We would like to investigate under which conditions we can 
factor from this functional all gauge transformations of $K$, leaving 
a functional integral over gauge transformations of $G$ modulo the ones of $K$.
In order to establish this factorisation, the following property is essential:
For fixed $A$ the quantity
\bq
 S\left[A,U\right] & = & - \frac{v^2}{4} \int || P_{\mathfrak b}(A^U) ||
\eq
is invariant under gauge transformations $V$ of $K$:
\bq
\label{invariance_H}
S\left[A, U V\right] & = & S\left[A,U\right].
\eq
To prove eq.~(\ref{invariance_H}) we first note
\bq
 A^{U V} & = & \left( A^U \right)^V 
 = V^{-1} \left( A^U \right) V + V^{-1} d V.
\eq
The term $V^{-1} d V$ is mapped to zero under $P_{\mathfrak b}$.
With the decomposition
\bq
 A^U & = & A^U_{\mathfrak k} + A^U_{\mathfrak h} + A^U_{\mathfrak e},
\;\;\;\;\;\;\;\;\;
 A^U_{\mathfrak k} \in {\mathfrak k},
 \;\;\;
 A^U_{\mathfrak h} \in {\mathfrak h},
 \;\;\;
 A^U_{\mathfrak e} \in {\mathfrak e},
\eq
we have
\bq
 V^{-1} \left( A^U_{\mathfrak k} + A^U_{\mathfrak h} \right) V & = & A^U_{\mathfrak k} + A^U_{\mathfrak h},
\eq
since $V$, $A^U_{\mathfrak k}$ and $A^U_{\mathfrak h}$ contain only the mutually commuting generators $H^a$.
For $A^U_{\mathfrak e}$ let us focus on a pair of generators $E^+$ and $E^-$, corresponding to root vectors
$\alpha$ and $-\alpha$.
We have
\bq
 \left[ H^a, E^+ \right] = \alpha_a E^+,
 & &
 \left[ H^a, E^- \right] = -\alpha_a E^-,
\eq
and using the Baker-Campbell-Hausdorff formula
we find
\bq
 e^{-i H^a \theta^a} E^+ e^{i H^a \theta^a} = e^{-i \theta^a \alpha_a} E^+,
 & &
 e^{-i H^a \theta^a} E^- e^{i H^a \theta^a} = e^{i \theta^a \alpha_a} E^-.
\eq
The term $||P_{\mathfrak b}(A^U_{\mathfrak e})||$ involves
\bq
 \mbox{Tr}\;\left( E^+ E^- \right).
\eq
Under the gauge transformation $V$ this term is transformed into
\bq
 \mbox{Tr}\;\left( e^{-i \theta^a \alpha_a} E^+ e^{i \theta^a \alpha_a} E^- \right)
 & = &
 \mbox{Tr}\;\left( E^+ E^- \right).
\eq
This completes the proof of eq.~(\ref{invariance_H}).
Note that for the proof we used the fact that $K$ is generated only by generators $H^a$.
In other words, it is required that $K$ is Abelian.
Let us summarise what we have established so far: We defined the functionals
\bq
\label{def_summary}
 Z[A] & = & \int {\cal D}U \; \exp\left( i S[A,U] \right),
 \nonumber \\
 S[A,U] & = & - \frac{v^2}{4} \int || P_{\mathfrak b}(A^U) ||,
\eq
with the following properties
\bq
 Z\left[A^{U_1}\right] & = & Z[A],
 \nonumber \\
 S\left[A, U V\right] & = & S\left[A,U\right],
\eq
where $U$ and $U_1$ denote gauge transformations of $G$ and $V$ denotes a gauge transformation of $K$.
The first equation states that $Z[A]$ is gauge invariant under all gauge transformation of $G$, 
the second equation states that for fixed $A$ the quantity $S[A,U]$ is invariant under
gauge transformations of $K$.

\subsection{Factorisation for infinitesimal gauge transformations}
\label{subsect:infinitessimal}

In this section let us assume that the functional integration ${\cal D}U$
in eq.~(\ref{def_Z3}) is restricted to infinitesimal gauge transformations.
For infinitesimal gauge transformations we can write a gauge transformation $U$ of $G$ as
\bq
 U & = & W V,
\eq
where $V$ is a gauge transformation of $K$ and $W$ is defined by
\bq
 W & = & 
 \exp\left( \chi \right),
\;\;\;
 \chi = i \sum\limits_{j=n+1}^N T^j \chi^j.
\eq
Note that the summation is only over the broken generators.
$W$ is a representative for the coset of gauge transformations of $G$ modulo the ones of $K$.
For infinitesimal transformations we have for the measure of $W$
\bq
 {\cal D}W & = & \prod\limits_{j=n+1}^N {\cal D}\chi^j,
\eq
and the measure ${\cal D}U$ factorises:
\bq
\label{factor_measure}
 {\cal D}U & = & {\cal D}\chi \cdot {\cal D}V.
\eq
Due to eq.~(\ref{invariance_H}) the integral over ${\cal D}V$ factorises from eq.~(\ref{def_Z3}):
\bq
 Z[A] & = & 
 \left( \int {\cal D}V \right) \cdot
 Z'[A],
\eq
with
\bq
\label{infinitessimal_result}
 Z'[A] & = & 
 \int {\cal D}W \exp\left( - i \frac{v^2}{4} \int || P_{\mathfrak b}(A^W) || \right).
\eq
The functional $Z'[A]$ has one free parameter $v$ and introduces 
for each broken generator $T^j$ a pseudo-Goldstone field $\chi^j$.
We derived eq.~(\ref{infinitessimal_result}) under the assumption
that all gauge transformations occurring in eq.~(\ref{def_Z3}) are infinitesimal.
Since eq.~(\ref{def_Z3}) involves an integration over all gauge transformation and not just infinitesimal ones, this 
assumption is of course not justified and I will discuss the corrections to eq.~(\ref{infinitessimal_result}) 
in the next section.
Nevertheless the result~(\ref{infinitessimal_result}) makes it transparent how the pseudo-Goldstone
fields emerge in this context: The pseudo-Goldstone fields are just the left-over fields, which cannot
be factorised from a functional involving an integration over all gauge transformations.

\subsection{Finite gauge transformations}
\label{subsect:finite}

In this section I discuss the modifications due to finite gauge transformations.
These modifications are entirely due to the measure of the integration.
In fact we have shown in eq.~(\ref{invariance_H}) that the integrand is invariant
under finite gauge transformations of $K$.
Let us denote the invariant measure of $U$ by ${\cal D}U$, the invariant measure of $V$ by
${\cal D}V$.
We parameterise a finite gauge transformation as
\bq
\label{def_U}
 U & = & 
 \exp\left( \chi \right),
\;\;\;
 \chi = i \sum\limits_{a=1}^N T^a \chi^a,
\eq
where the sum is now over all generators of $G$.
The measure ${\cal D}U$ expressed in the fields $\chi^a$ is
\bq
 {\cal D}U 
 & = & 
 \prod\limits_{a=1}^N {\cal D}\chi^a \; \mbox{det}\; M_G(\chi).
\eq
At a fixed space-time point this is just the Haar measure of the group $G$.
The determinant $\mbox{det}\; M_G(\chi)$ can be obtained as follows: Let us define a matrix
$N_G$ through
\bq
 N_G^{ab} & = & i f^{acb} \chi^c.
\eq
Then $M_G$ is given by
\bq
\label{def_M}
 M_G & = & \sum\limits_{n=0}^\infty \frac{1}{(n+1)!} \left( - i N_G \right)^n.
\eq
For $G=SU(2)$ the determinant can actually be calculated explicitly.
If we use as generators $T^a = \frac{1}{2} \sigma^a$, where $\sigma^a$ are the Pauli matrices,
one finds
\bq
\label{det_explicit}
\mbox{det}\;M_G(\chi) & = & \left( \frac{\sin \frac{\rho}{2}}{\frac{\rho}{2}} \right)^2,
\;\;\;\;\;\;
\rho = \sqrt{ \left(\chi^1\right)^2 + \left(\chi^2\right)^2 + \left(\chi^3\right)^2}.
\eq
For $G=U(1)$ the measure is constant:
\bq
\mbox{det}\;M_G(\chi) & = & 1.
\eq
Let us now denote by ${\cal D}W$ the $G$-invariant measure of the coset space $G/K$.
This measure satisfies
\bq
 \int {\cal D}U f(U) & = &
 \int {\cal D}W \int {\cal D}V f(WV)
\eq
for any function $f$ defined on $G$.
If the function $f$ is invariant under transformations of $K$, e.g. $f(WV)=f(W)$, we obtain
\bq
 \int {\cal D}U f(U) & = &
 \left( \int {\cal D}V  \right) \int {\cal D}W f(W).
\eq
In the case of interest here we have $f(U) = \exp(i S[A,U])$ and the function is invariant
under transformations of $K$.
We define
\bq
\label{def_J}
 {\cal J} & = & \int {\cal D}V \; \mbox{det}\; M_G.
\eq
Then the measure ${\cal D}W$ is given up to normalisation factors by
\bq
 {\cal D}W & = & \prod\limits_{j=n+1}^N {\cal D}\chi^j \; {\cal J}.
\eq
Putting everything together we arrive at the final formula for the functional $Z'[A]$:
\bq
\label{final_formula}
 Z'[A] & = & 
 \int \prod\limits_{j=n+1}^N {\cal D}\chi^j \; {\cal J} \;  \exp\left( - i \frac{v^2}{4} \int || P_{\mathfrak b}(A^W) || \right).
 \nonumber \\
\eq
Formula~(\ref{final_formula}) is the main result of this letter.
Note that we can also write for $A^W$
\bq
\label{def_covariant_derivative}
 A^W & = & W^{-1} A W + W^{-1} d W = W^{-1} D W,
\eq
where $D=d+A$ is the covariant derivative.
We can therefore write
\bq
 - \frac{v^2}{4} \int || P_{\mathfrak b}(A^W) ||
 & = & 
 \frac{v^2}{4} \int d^4x \; 
 \mbox{Tr} \; \left( P_{\mathfrak b}\left(\left( D_\mu W \right)^\dagger W \right) P_{\mathfrak b}\left(W^{-1} D^\mu W\right) \right).
\eq
We now define ${\cal L}_{breaking}$ by
\bq
\label{Lagrangian_breaking}
 {\cal L}_{breaking} 
 & = & 
 \frac{v^2}{4} \mbox{Tr} \; \left( P_{\mathfrak b}\left(\left( D_\mu W \right)^\dagger W \right) P_{\mathfrak b}\left(W^{-1} D^\mu W\right) \right).
\eq
Then
\bq
 Z'[A] & = &
 \int \prod\limits_{j=n+1}^N {\cal D}\chi^j \; {\cal J} \;
 \exp\left( i \int d^4x \; {\cal L}_{breaking}
     \right).
 \;\;
\eq
We can expand ${\cal L}_{breaking}$ in the pseudo-Goldstone fields
\bq
 {\cal L}_{breaking} & = & \sum\limits_{n=0}^\infty {\cal L}^{(n)}_{breaking},
\eq
such that ${\cal L}^{(n)}_{breaking}$ contains $n$ pseudo-Goldstone fields.
The first term is given by
\bq
 {\cal L}^{(0)}_{breaking}
 & = & 
 \frac{v^2 g^2}{8} \sum\limits_{a=n+1}^N \left( A^a_\mu \right)^\dagger A^{a \mu}.
\eq

\section{Application to the electro-weak theory}
\label{sect:application}

In this section I apply the results of the previous section to an $SU_L(2) \times U_Y(1)$ gauge symmetry.
I first discuss the masses of the gauge bosons in section~\ref{subsect:masses}.
The $\rho$-parameter is discussed in section~\ref{subsect:rho-parameter}.
The equivalence with the standard lowest order effective Lagrangian is shown in section~\ref{subsect:effective}.
The effects of the measure are discussed in section~\ref{subsect:measure}.

\subsection{The masses of the gauge bosons}
\label{subsect:masses}

We start from the Lagrange density
\bq
{\cal L}_{gauge} & = & 
 - \frac{1}{4} W_{\mu\nu}^a W^{\mu\nu a}
 - \frac{1}{4} B_{\mu\nu} B^{\mu\nu},
\eq
where
\bq
W^a_{\mu\nu} & = & \partial_\mu W^a_\nu - \partial_\nu W^a_\mu 
 + g f^{abc} W^b_\mu W^c_\nu, \nonumber \\
B_{\mu\nu} & = & \partial_\mu B_\nu - \partial_\nu B_\mu.
\eq
$W^a_{\mu\nu}$ is the field strength corresponding to $SU_L(2)$, $B_{\mu\nu}$ is the field strength 
corresponding to $U_Y(1)$.
The covariant derivative is
\bq
D_\mu& = & \partial_\mu - i g I^a W^a_\mu - i g' I^0 B_\mu,
\eq
where $I^a = \frac{1}{2} \sigma^a$ for $a\in \{1,2,3\}$ ($\sigma^a$ are the Pauli matrices) and
$I^0 = \frac{1}{2} {\bf 1}$.
The coupling constant of $SU_L(2)$ is denoted by $g$, the one of $U_Y(1)$ is denoted by $g'$.
The generators are normalised as
\bq
\label{normalisation_ew}
 \mbox{Tr}\; I^a I^b & = & \frac{1}{2} \delta^{ab}.
\eq
Let us now define
\bq
\label{def_generators}
 H^0 = \frac{1}{\sqrt{2}} \left( I^0 + I^3 \right),
 \;\;\;
 H^1 = I^3,
 \;\;\;
 E^\pm = \frac{1}{\sqrt{2}} \left( I^1 \pm i I^2 \right).
\eq
The set $\{H^0, H^1, E^+, E^- \}$ defines another base.
Note that $H^0$ and $H^1$ are not orthogonal.
We now consider the case, where the symmetry group $SU_L(2) \times U_Y(1)$ is broken down to a symmetry group
$U_{QED}(1)$ generated by $H^0$.
The gauge potential can be written as
\bq
\label{gauge_potential}
 g I^a W^a_\mu + g' I^0 B_\mu
 & = &
 g W_\mu^+ E^+ + g W_\mu^- E^-
 + \left( g W_\mu^3 - g' B_\mu \right) H^1
 + \sqrt{2} g' B_\mu H^0,
\eq
where we set
\bq
W^\pm_\mu & = & \frac{1}{\sqrt{2}} \left( W^1_\mu \mp i W^2_\mu \right).
\eq
We find for the projection 
\bq
 P_{\mathfrak b}(A)  
 & = &
 \frac{1}{i} \left( g W^+_\mu E^+ + g W^-_\mu E^-
                  + \left( g W^3_\mu - g' B_\mu \right) H^1 \right) dx^\mu
\eq
and therefore
\bq
\label{Lagrangian_mass}
{\cal L}^{(0)}_{breaking} 
 & = &
 \frac{v^2}{8} 
 \left[
  2 g^2 W^+_\mu W^{-\mu}
  + \left( B_\mu, W^3_\mu \right)
    \left( \begin{array}{cc}
           {g'}^2 & - g g' \\
           -g g' & g^2 \\
           \end{array} \right)
    \left( \begin{array}{c} B^\mu \\ W^{3\mu} \\ \end{array} \right)
 \right].
\eq
The matrix is diagonalised as usual by
\bq
\left( \begin{array}{c} A_\mu \\ Z_\mu \\ \end{array} \right)
& = & 
\left( \begin{array}{cc} \cos \theta_W & \sin \theta_W \\
- \sin \theta_W & \cos \theta_W \\ \end{array} \right)
\left( \begin{array}{c} B_\mu \\ W^3_\mu \\ \end{array} \right)
\eq
with
\bq
\cos \theta_W = \frac{g}{\sqrt{g^2+{g'}^2}}, & &
\sin \theta_W = \frac{g'}{\sqrt{g^2+{g'}^2}}.
\eq
One obtains the standard mass term
\bq
\label{Lagrangian_mass_final}
{\cal L}^{(0)}_{breaking} 
 & = & 
 m_W^2 W^+_\mu W^{-\mu}
 + \frac{1}{2} m_Z^2 Z_\mu Z^\mu,
\eq
with
\bq
m_W = \frac{v}{2} g, & & 
m_Z = \frac{v}{2} \sqrt{g^2 + {g'}^2}.
\eq
The symmetry breaking parameter $v$ can be expressed in terms of measured quantities
as
\bq 
 v & = & \frac{2 \sin \theta_W}{e} m_W,
\eq
where $e = g g' / \sqrt{g^2+{g'}^2}$ is the electric charge.

\subsection{The $\rho$-parameter}
\label{subsect:rho-parameter}

In the previous section we made a choice for the generator $H^0$ as a base for $\mathfrak k$
and a choice for the generators $\{H^1,E^+,E^-\}$ as a base for $\mathfrak b$.
The choice of $H^0$ is motivated by the quantum numbers of the fermions
and the choice of $E^\pm$ is also rather un-problematic.
However, the choice of $H^1$ deserves some discussion.
The specific choice $H^1=I^3$ makes the vector spaces $\mathfrak k$ and $\mathfrak b$ non-orthogonal, and raises
the question why the choice
\bq
\label{orthogonal_choice}
 H^1 & = & \frac{1}{\sqrt{2}} \left( I^0 - I^3 \right),
\eq
which would ensure orthogonality, is not used.
To investigate this question let us assume that $H^1$ is a linear combination of $I^0$ and $I^3$:
\bq
 H^1 & = & s I^0 + c I^3.
\eq
The normalisation $\mbox{Tr} H^1 H^1 = 1/2$ requires
\bq
 s^2 + c^2 & = & 1,
\eq
therefore $s$ and $c$ are the sine and cosine of some angle.
We can repeat the analysis of the previous section and find that
the mass of the $Z$-boson is now
\bq
 m_Z^2 & = & \frac{v^2}{4} \frac{g^2+{g'}^2}{(s+c)^2}
\eq
One then finds for the $\rho$-parameter
\bq
 \rho & = & \frac{m_W^2}{m_Z^2 \cos^2 \theta_W}
 =
 1 + 2 s c.
\eq
If we require that at tree-level we have $\rho=1$, it follows that $s=0$ or $c=0$.
Therefore the allowed choices for $H^1$ are $\pm I^3$ and $\pm I^0$.
This excludes the orthogonal choice in eq.~(\ref{orthogonal_choice}).
 
\subsection{The effective Lagrangian}
\label{subsect:effective}

In this section I show that for the breaking of the electro-weak symmetry 
the Lagrangian ${\cal L}_{breaking}$ given in eq.~(\ref{Lagrangian_breaking})
reduces to the lowest-order Lagrangian for chiral symmetry breaking 
given in eq.~(\ref{Lagrangian_chiralSB}).
To this aim let us consider the pseudo-Goldstone fields in the electro-weak theory.
The coset space, defined through the generators $\{H^1, E^+, E^- \}$, is in this case actually 
a group.
As can be seen from eq.~(\ref{def_generators}), $H^1$, $E^+$ and $E^-$ are the generators of $SU(2)$.
A general element of the coset space can be parameterised as
\bq
\label{coset_representative1}
 \exp \left( i \left( H^1 \chi + E^+ \phi^+ + E^- \phi^- \right) \right) 
         \exp \left( i H^0 \xi \right).
\eq
We will choose as coset representative $\xi = 0$ and since the coset space is 
a group we change notation and label the coset representative by U:
\bq
\label{coset_representative2}
 U & = & \exp \left( i \left( H^1 \chi + E^+ \phi^+ + E^- \phi^- \right) \right).
\eq
A $SU_L(2) \times U_Y(1)$ gauge transformation acts on $U$ as follows:
\bq
\label{Goldstone_transformation}
 \exp \left( i I^a \theta^a + i I^0 \theta^0\right) U 
 & = & 
 U' \exp \left( i \sqrt{2} H^0 \theta^0 \right),
\eq
with 
\bq
\label{Goldstone_transformation2}
 U' & = & \exp \left( i I^a \theta^a \right) U \exp \left( - i I^3 \theta^0 \right).
\eq
Note that $U'$ is again of the form as in eq.~(\ref{coset_representative2}).
To derive eq.~(\ref{Goldstone_transformation}) we used the fact that $I^0$ commutes
with all generators.
The covariant derivative acts on $U$ as follows
\bq
\label{covariant_derivative}
 D_\mu U & = &
 \partial_\mu U
 - i g I^a W^a_\mu U
 + i g' U I^3 B_\mu.
\eq
Note that the fact that $U_Y(1)$ acts through $I^3$ from the right is a consequence
of eq.~(\ref{Goldstone_transformation2}).
As eq.~(\ref{covariant_derivative}) involves only the generators of $SU(2)$, but not $I^0$, we may
drop the projection $P_{\mathfrak b}$ in ${\cal L}_{breaking}$.
We obtain
\bq
 {\cal L}_{breaking} & = & 
 \frac{v^2}{4} \mbox{Tr} \; \left( \left( D_\mu U \right)^\dagger D^\mu U \right).
\eq
This is the usual lowest-order effective Lagrangian for electro-weak symmetry breaking.

\subsection{Effects of the measure}
\label{subsect:measure}

In the formula eq.~(\ref{final_formula}) we had the additional factor ${\cal J}$ related 
to the measure in the path integral.
In the example of electro-weak theory, the integration over ${\cal D}V$ in eq.~(\ref{def_J}) can be factored, leaving
just the determinant related to the Haar measure of $SU_L(2)$.
With the parameterisation as in eq.~(\ref{coset_representative2}) for the coset space
we find that the matrix $N_G$ is given in the base $(\chi, \phi^+, \phi^-)^T$ by
\bq
 N_G & = & 
 \left( \begin{array}{ccc}
 0 & \phi^- & - \phi^+ \\
 \phi^+ & - \chi & 0 \\
 -\phi^- & 0 & \chi \\
 \end{array} \right)
\eq
and
\bq 
 {\cal J} & = & \mbox{det}\; M_G,
 \;\;\;
 M_G = \sum\limits_{n=0}^\infty \frac{\left( - i N_G \right)^n}{(n+1)!}.
\eq
The determinant can be exponentiated with the Faddeev-Popov method \cite{Faddeev:1967fc}:
\bq
 {\cal J}
 & = & 
 \int {\cal D}c {\cal D}\bar{c} \;
 \exp \left( i \int d^4x \; \bar{c} \; M_G \; c \right).
\eq
$\bar{c}=(\bar{c}_0,\bar{c}_-,\bar{c}_+)$ and $c=(c_0,c_+,c_-)^T$ are the Faddeev-Popov ghosts for the pseudo-Goldstone fields $\chi$ and $\phi^\pm$.
They are Grassmann-valued fields and in this specific case they have the
peculiar property that they are non-propagating fields in the sense
that their ``propagator'' is simply $i$.
The net effect of these ghosts in loops consists in generating contact interactions
among the pseudo-Goldstone fields, which are multiplied by quartic divergent integrals
\bq
 \int \frac{d^4k}{(2\pi)^4} \cdot 1.
\eq
In dimensional regularisation these integrals are zero and the effects of the measure can be ignored.

Note that in this letter we just discuss the lowest-order effective Lagrangian containing two derivatives.
Of course this Lagrangian should be supplemented with terms containing a higher number of derivatives.
The renormalisation of the effective theory is similar to the case of chiral perturbation theory \cite{Gasser:1983yg,Gasser:1984gg}.

\subsection{Discussion}
\label{subsect:discussion}

Let us summarise the essential points:
\begin{itemize}
\item In order to derive the lowest order effective Lagrangian for electro-weak symmetry breaking
with a value $\rho=1$ of the $\rho$-parameter
we assumed a projection $P_{\mathfrak b}$ onto the broken generators. 
The kernel of this projection is given by the un-broken generators.
For the electro-weak theory with an $SU_L(2) \times U_Y(1)$ gauge group $P_{\mathfrak b}$ projects onto
$SU_L(2)$. The kernel is spanned by the generator of $U_{QED}(1)$.
This ensures $\rho=1$.
Note that this projection gives rise to a decomposition of the Lie algebra of $SU_L(2) \times U_Y(1)$ as a vector
space, which is not orthogonal with respect to the standard inner product.

In the formalism presented here this projection plays the role the custodial symmetry $SU_{L+R}(2)$ plays
in the approach of chiral electro-weak symmetry breaking.

\item The pseudo-Goldstone fields are the left-over fields, which cannot
be factorised from a functional involving an integration over all gauge transformations.
They parameterise the coset space $G/K$. 
For the electro-weak theory with the decomposition as above the coset space is $SU_L(2)$, therefore the 
pseudo-Goldstone fields are directly associated to $SU_L(2)$.

In contrast, in chiral electro-weak symmetry breaking they are associated with the spontaneous breaking
of a global $SU_{L-R}(2)$ symmetry.

\item The fact that in the covariant derivative eq.~(\ref{covariant_derivative})
the field $B_\mu$ acts through $I^3$ from the right is a consequence of the transformation properties
of the coset representative under gauge transformations of $SU_L(2)\times U_Y(1)$.
\end{itemize}

\section{Conclusions}
\label{sect:conclusions}

In this letter I constructed a gauge-invariant functional for the symmetry breaking of a gauge group $G$
down to an Abelian sub-group $K$.
This functional involves an integration over all gauge transformation of $G$.
The gauge transformations of $K$ can be factored out.
The pseudo-Goldstone modes emerge naturally from the fact that it is impossible to factorise the remaining integration
over gauge transformations related to the broken generators.
Applied to electro-weak theory this functional coincides with the standard lowest order effective Lagrangian
for electro-weak symmetry breaking.
The important point is that in the approach presented here this effective Lagrangian is derived with
minimal assumptions: Neither was first the spontaneous breaking of a global symmetry assumed nor was an additional
custodial symmetry assumed.


\bibliography{/home/stefanw/notes/biblio}

\begin{thebibliography}{10}

\bibitem{Higgs:1964ia}
P.~W. Higgs,
\newblock Phys. Lett. {\bf 12}, 132 (1964).

\bibitem{Higgs:1964pj}
P.~W. Higgs,
\newblock Phys. Rev. Lett. {\bf 13}, 508 (1964).

\bibitem{Higgs:1966ev}
P.~W. Higgs,
\newblock Phys. Rev. {\bf 145}, 1156 (1966).

\bibitem{Englert:1964et}
F.~Englert and R.~Brout,
\newblock Phys. Rev. Lett. {\bf 13}, 321 (1964).

\bibitem{Guralnik:1964eu}
G.~S. Guralnik, C.~R. Hagen, and T.~W.~B. Kibble,
\newblock Phys. Rev. Lett. {\bf 13}, 585 (1964).

\bibitem{Kibble:1967sv}
T.~W.~B. Kibble,
\newblock Phys. Rev. {\bf 155}, 1554 (1967).

\bibitem{Goldstone:1961eq}
J.~Goldstone,
\newblock Nuovo Cim. {\bf 19}, 154 (1961).

\bibitem{Goldstone:1962es}
J.~Goldstone, A.~Salam, and S.~Weinberg,
\newblock Phys. Rev. {\bf 127}, 965 (1962).

\bibitem{Cornwall:1974km}
J.~M. Cornwall, D.~N. Levin, and G.~Tiktopoulos,
\newblock Phys. Rev. {\bf D10}, 1145 (1974).

\bibitem{Lee:1977eg}
B.~W. Lee, C.~Quigg, and H.~B. Thacker,
\newblock Phys. Rev. {\bf D16}, 1519 (1977).

\bibitem{Chanowitz:1985hj}
M.~S. Chanowitz and M.~K. Gaillard,
\newblock Nucl. Phys. {\bf B261}, 379 (1985).

\bibitem{Chanowitz:1986hu}
M.~S. Chanowitz, M.~Golden, and H.~Georgi,
\newblock Phys. Rev. Lett. {\bf 57}, 2344 (1986).

\bibitem{Chanowitz:1987vj}
M.~S. Chanowitz, M.~Golden, and H.~Georgi,
\newblock Phys. Rev. {\bf D36}, 1490 (1987).

\bibitem{Appelquist:1980vg}
T.~Appelquist and C.~W. Bernard,
\newblock Phys. Rev. {\bf D22}, 200 (1980).

\bibitem{Longhitano:1980iz}
A.~C. Longhitano,
\newblock Phys. Rev. {\bf D22}, 1166 (1980).

\bibitem{Longhitano:1980tm}
A.~C. Longhitano,
\newblock Nucl. Phys. {\bf B188}, 118 (1981).

\bibitem{Donoghue:1989qw}
J.~F. Donoghue and C.~Ramirez,
\newblock Phys. Lett. {\bf B234}, 361 (1990).

\bibitem{Dawson:1990cc}
S.~Dawson and G.~Valencia,
\newblock Nucl. Phys. {\bf B352}, 27 (1991).

\bibitem{Dobado:1989ax}
A.~Dobado and M.~J. Herrero,
\newblock Phys. Lett. {\bf B228}, 495 (1989).

\bibitem{Dobado:1990zh}
A.~Dobado, D.~Espriu, and M.~J. Herrero,
\newblock Phys. Lett. {\bf B255}, 405 (1991).

\bibitem{Espriu:1991vm}
D.~Espriu and M.~J. Herrero,
\newblock Nucl. Phys. {\bf B373}, 117 (1992).

\bibitem{Sikivie:1980hm}
P.~Sikivie, L.~Susskind, M.~B. Voloshin, and V.~I. Zakharov,
\newblock Nucl. Phys. {\bf B173}, 189 (1980).

\bibitem{Veltman:1976rt}
M.~J.~G. Veltman,
\newblock Acta Phys. Polon. {\bf B8}, 475 (1977).

\bibitem{Veltman:1977kh}
M.~J.~G. Veltman,
\newblock Nucl. Phys. {\bf B123}, 89 (1977).

\bibitem{Coleman:1969sm}
S.~R. Coleman, J.~Wess, and B.~Zumino,
\newblock Phys. Rev. {\bf 177}, 2239 (1969).

\bibitem{Faddeev:1967fc}
L.~D. Faddeev and V.~N. Popov,
\newblock Phys. Lett. {\bf B25}, 29 (1967).

\bibitem{Gasser:1983yg}
J.~Gasser and H.~Leutwyler,
\newblock Ann. Phys. {\bf 158}, 142 (1984).

\bibitem{Gasser:1984gg}
J.~Gasser and H.~Leutwyler,
\newblock Nucl. Phys. {\bf B250}, 465 (1985).

\end{thebibliography}
\bibliographystyle{/home/stefanw/latex-style/h-physrev3}

\end{document}